\documentclass[aps,prl,reprint,twocolumn,preprintnumbers,floatfix,nofootinbib]{revtex4-1}

\usepackage[T1]{fontenc} 

\usepackage[dvipdfm]{graphicx}
\usepackage{epstopdf}
\usepackage{mathrsfs}
\usepackage{amssymb}
\usepackage{verbatim}
\usepackage{color}
\usepackage[dvipsnames]{xcolor}
\usepackage{multirow}
\usepackage{amsmath}
\usepackage{subfig}
 \usepackage{slashed} 
\usepackage{float}
\usepackage[normalem]{ulem}
\usepackage{sidecap}
\usepackage{hyperref}
\usepackage{cancel}
\usepackage{enumerate}
\hypersetup{pdfstartview=FitV,colorlinks=true,linkcolor=blue,citecolor=red,filecolor=black,urlcolor=blue}

\newcommand*\diff{\mathop{}\!\mathrm{d}}

\newcommand{\beq}{\begin{equation}}
\newcommand{\eeq}{\end{equation}}
\newcommand{\bea}{\begin{eqnarray}}
\newcommand{\eea}{\end{eqnarray}}

\newcommand{\mrm}{\mathrm}

\def\to{\rightarrow}

\usepackage[parfill]{parskip}
\begin{document}
\sloppy 

\preprint{UWThPh 2020-16}
\preprint{IFT-UAM/CSIC-20-107}
\preprint{UMN--TH--3921/20, FTPI--MINN--20/20}

\vspace*{1mm}

\title{Energy--Momentum portal to dark matter
and emergent gravity} 

\author{Pascal Anastasopoulos$^{a}$}
\email{pascal.anastasopoulos@univie.ac.at}
\author{Kunio Kaneta$^{b}$}
\email{kkaneta@umn.edu}
\author{Yann Mambrini$^{c}$}
\email{yann.mambrini@ijclab.in2p3.fr}
\author{Mathias Pierre$^{d,e}$}
\email{mathias.pierre@uam.es}
\vspace{0.5cm}

\affiliation{$^a$Mathematical Physics Group, University of Vienna, Boltzmanngasse 5 1090 Vienna, Austria}

\affiliation{$^b$William I. Fine Theoretical Physics Institute, School of
 Physics and Astronomy, University of Minnesota, Minneapolis, MN 55455,
 USA}

\affiliation{$^c$Universit\'e Paris-Saclay, CNRS/IN2P3, IJCLab, 91405 Orsay, France}

 \affiliation{$^d$Instituto de F\'{i}sica Te\'{o}rica (IFT) UAM-CSIC, Campus de Cantoblanco, 28049 Madrid, Spain} 
\affiliation{
$^e$
Departamento de F\'{i}sica Te\'{o}rica, Universidad Aut\'{o}noma de Madrid (UAM), Campus de Cantoblanco, 28049 Madrid, Spain}

\date{\today}

\begin{abstract} 
 We propose a new scenario where dark matter belongs to a secluded sector coupled to the Standard Model through energy--momentum tensors. 
 Our model is motivated by constructions where gravity {\it emerges} from a hidden sector,
the graviton being identified by the kinetic term of the fields in the secluded sector. 
Supposing that the lighter particle of the secluded sector is the dark component of the Universe, we show that we can produce it in a sufficiently large amount despite the suppressed couplings of the theory, thanks to large temperatures of the thermal bath in the early stage of the Universe. 

\vskip 1cm

\begin{center}
{\it \large Dedicated to the memory of Renaud Parentani}
\end{center}

\vskip 1cm

\end{abstract}

\maketitle

\setcounter{equation}{0}

\section{I. Introduction}

Almost 90 years ago, in 1933, F. Zwicky in a seminal paper~\cite{Zwicky:1933gu} observed an anomaly in the virial velocities of the galaxies forming the Coma cluster. Two years later, a young PhD student, H. Babcock~\cite{babcock}, unaware of the Zwicky's work noticed a strange behavior in the rotation curve of Andromeda. Both reached the same conclusion: the need for an additional  (dark) component in our Universe. Several studies, especially concerning the stability of large scale structures~\cite{Ostriker:1973uit} confirmed this hypothesis, until the proposal of the addition of a new weakly interacting massive neutral particle (WIMP) by Steigman {\it et al.}~\cite{Gunn:1978gr} in 1978.

The main elegance of the proposal was the complete independence of the relic abundance on the thermal history of the Universe: as long as we suppose that, anytime before the decoupling, the WIMP $\chi$ was in equilibrium with the bath, the relic abundance is {\it only} dependent on its mass and the coupling between the dark and visible sectors. However, the last experiments dedicated to direct detection measurements exclude proton-WIMP cross section $\sigma_{\chi-p} \gtrsim 10^{-46} \rm{cm^2}$  for a 100 GeV dark matter candidate~\cite{XENON,LUX,PANDAX}, which is six orders of magnitude below the vanilla
Fermi cross section for weakly interacting particle~\cite{Higgsportal,Zportal}. Any simple Beyond-the-Standard-Model (BSM) extension which can reproduce the relic abundance observed by Planck~\cite{planck} and avoid such strong constraints, without relying on some corner of parameter space, requires to invoke a new physics scale $\simeq 5$ TeV~\cite{Arcadi:2017kky}
The next generation of direct detection experiments will probe cross sections as weak as $\sigma_{\chi-p}\gtrsim 10^{-49} \rm{cm^2}$~\cite{Aalbers:2016jon}. If no signal is seen, that would push the BSM scale above 50 TeV, much above the electroweak scale breaking\footnote{Notice that recently, XENON1T experiment proved that light sector of dark matter can be probed through electron recoil \cite{electronXENON1T}.}. 

The conflict existing between direct detection observations and the relic abundance constraints lies mainly on the fact that we supposed that dark matter was in thermal equilibrium with the Standard Model (SM) sector before decoupling from it. Relaxing this hypothesis opens new windows on dark matter phenomenology, especially
if one considers FIMP (Freeze In Massive Particle) candidates~\cite{fimp,Bernal:2017kxu}.
In such scenario, the dark sector is highly secluded 
from the SM by a feeble coupling or by a very massive mediators\footnote{We then discuss UV freeze in~\cite{Elahi:2014fsa}.} which can be a $Z'$ \cite{Bhattacharyya:2018evo}, moduli fields~\cite{Chowdhury:2018tzw}, massive spin-2 particles~\cite{Bernal:2018qlk}, in the SO(10) framework~\cite{SO10}. Even mediators belonging to the inflaton sector itself~\cite{HighlyDecoupled}, or a Kaluza-Klein framework~\cite{Garny:2017kha,Bernal:2020fvw} and spin-$\frac{3}{2}$ particles \cite{Garcia:2020hyo} has been recently analyzed. The case of supergravity is slightly different.
Indeed, whereas in low-scale supersymmetry (SUSY) its production rate is suppressed
by Planck-mass coupling~\cite{gravitino}, in high-scale SUSY, when the spectrum is above the inflaton mass~\cite{Benakli:2017whb}, the temperature dependence renders the dark matter production very sensible to the early stages of reheating. In all these cases, it was shown that, due to the high dependence of
the scattering rates on the temperature (because of mass-suppressed processes), 
the influence of the early Universe physics is strong, especially during the reheating phase~\cite{Giudice:2000ex,Garcia:2017tuj,Elahi:2014fsa,Reheating}. Moreover, the (strong) influence of the inflaton equation of state~\cite{Garcia:2020eof} or its decay modes~\cite{Kaneta:2019zgw} was added to recent studies.

On another hand, gravity is one of the biggest puzzles of physics. After years of attempts a quantum theory of gravity is still lacking, one of the main problem being its non-renormalizability. 
The closest approach to quantum gravity is coming from string theory which provides a well defined perturbative quantum gravity at semi-classical level. It introduces a scale, the string scale, which bypasses the renormalizability issues~\cite{Kiritsis:2007zza}.

At this point, AdS/CFT correspondence and the holographic ideas can provide a non-perturbative and UV-complete theory of quantum gravity. 
Recently, some authors adapted holographic ideas and suggested that the whole physics is described by four-dimensional quantum field theories~\cite{Kiritsis:2014yqa, BBKNpaper, AxionABBCK, globalcurrentstoU1, ABCKgraviphotons}. In such a framework, gravity is 
{\it not} a fundamental interaction but it emerges from a hidden sector which does not directly couple to the SM but only via very massive messenger fields.
In~\cite{Kiritsis:2014yqa, BBKNpaper} the authors study the emergence of gravity, where the graviton is considered as a composite field and is proportional to the energy-momentum tensor of the hidden theory. 
Below the messenger scale, this composite/emergent graviton behavior is described by the General Relativity action and the cosmological constant depends on various parameters of the hidden sector\footnote{
This model bypasses the Weinberg-Witten theorem since it predicts a non vanishing cosmological constant~\cite{BBKNpaper}.}.

AdS/CFT correspondence has taught us that there is a higher-dimensional picture of the framework described above.
In that picture, the SM sector is localized on a four-dimensional brane, which is immersed in a five-dimensional space-time at an appropriate radial direction corresponding to a messenger mass cutoff. The hidden sector lives in the bulk (all five dimensions). 

The fact that the whole theory (the SM, the hidden sector and the messenger sector) forms an UV-complete theory is very tempting for further study and especially at phenomenological level. Since gravity is the only interaction, known to this day, for which the dark matter component of the universe is sensitive to, constructions where gravity and dark matter both arise from a more complex sector is one of the most obvious and motivated possibilities. Indeed, it seems natural that 
the secluded sector from which emerges the gravity, contains as one of its components, if stable, the dark matter. The communication between the two sectors being highly suppressed when dealing with scattering processes below the inflaton mass, the FIMP paradigm is clearly justified {\it and} extremely natural, the strengths of the interaction being directly related to the masses in the messengers sectors.
The paper is organized as follows. After a description of our setup and the motivation lying beyond our construction in section II, we compute the production rate and dark matter abundance in section III, refining our analysis taking into account subtleties of reheating scenarios before concluding in section IV.






\section{II. The model}

\subsection{Motivations}

\begin{figure}[ht]
\centering
\includegraphics[width=2.3 in]{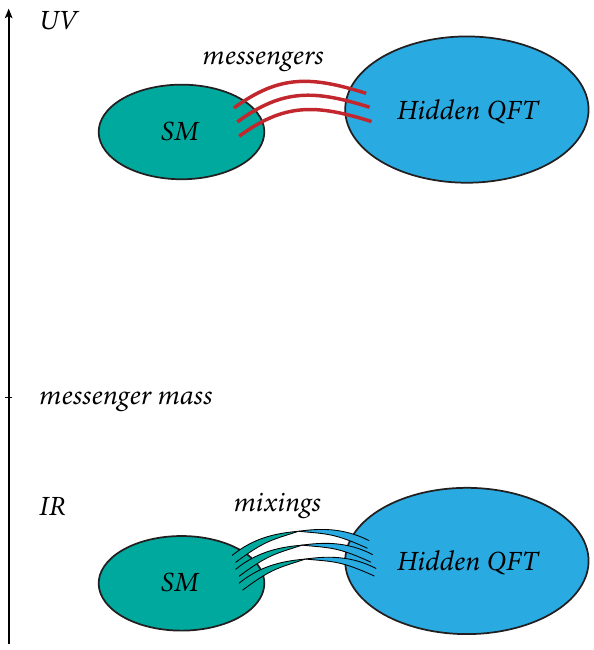}
\caption{\em \small Schematic picture of our setup}
\label{Fig:hidden}
\end{figure}

Following~\cite{Kiritsis:2014yqa, BBKNpaper, AxionABBCK, globalcurrentstoU1}, we will consider two secluded sectors, a visible one where live the SM fields respecting the $SU(3)_c\times SU(2)_L\times U(1)_Y$ (or even larger) gauge symmetries, and a hidden sector described by a hidden quantum field theory. The two sectors 
communicate only through the exchange of very massive fields. These messengers
are bifundamental fields, charged both under the SM gauge groups {\it and} the hidden gauge group(s). We suppose that their masses $M^i_\text{mess}$ are much above the reheating temperature, we illustrate our setup in Fig.~\ref{Fig:hidden}. 

The spectrum can be very rich, we will then by simplicity call
$\Lambda$ a common mass scale for all the messenger sector. 
At a scale much above $\Lambda$, the two sectors communicate quite efficiently. However, for processes occurring at an infrared scale below $\Lambda$, the two sectors
begin to be secluded. This is exactly what can happen just after inflation, 
where the reheating temperature generated by the inflaton decay, $T_{RH} \simeq \sqrt{\Gamma^\Phi~M_P}\simeq 10^{11}$ GeV ($\Gamma^\Phi$ being the width of the inflaton $\Phi$ and $M_P$ the Planck mass\footnote{We will use throughout our work $M_P=2.4 \times 10^{18}$ GeV for the reduced Planck mass.}) is much lower than the inflationary scale at the end of inflation, $\Phi_{end} \simeq {\cal O}(M_P)$. We will show that despite such a seclusion, the reheating temperature could be sufficiently large to produce the right amount of relic abundance, where 
the lightest stable particle in the hidden sector is a perfect candidate for dark matter. Despite the fact that its density has been completely diluted after the inflation, we will show that it can be repopulated during the reheating period from the thermal bath and inflaton decay, through a new type of portal. 



\subsection{The Lagrangian}

As we will see in the following, emergent type of gravity couples energy--momentum tensors of SM field to the energy--momentum tensors in the hidden sector, through the emergent metric generated by the mechanism. To avoid treating a specific model and to stay as general as possible, we decide first to study a generic new type of portal, where the dark and visible sectors interact through their own energy--momentum tensors. As we will see, other models like spin-2 portal, or high-scale supergravity, possesses similar characteristics when considered as effective theories. At first, we consider the specific type of effective coupling given by
\begin{equation}
    S_\text{portal}  =   \int \diff^4 x \sqrt{-g} \left( -\dfrac{1}{2 \Lambda^4} T_\text{SM}^{\mu \nu} T_{\mu \nu}
    \right)
    \label{eq:interaction_energymomentumtensor}
\end{equation}
with $T_\text{SM}^{\mu\nu}$ the energy--momentum tensor of the SM fields, and $T_{\mu \nu} = \sum_i T_{\mu \nu}^i$ the energy--momentum tensor of hidden fields which, depending on the spin $i$ of the concerned fields\footnote{We assume real scalars and Dirac fermions throughout our work.}, can be written
\bea
T^0_{\mu \nu} &=& \frac{1}{2} \left( \partial_\mu X~\partial_\nu X + \partial_\nu X ~\partial_\mu X -g_{\mu \nu} \partial^\alpha X ~\partial_\alpha X \right) 
\,, \nonumber\\
T^{1/2}_{\mu \nu} &=& \frac{i}{4}
\bar \chi \left( \gamma_\mu \partial_\nu + \gamma_\nu \partial_\mu \right) \chi
-\frac{i}{4} \left( \partial_\mu \bar \chi \gamma_\nu + \partial_\nu \bar \chi \gamma_\mu \right)\chi \,, 
\nonumber\\
T^{1}_{\mu \nu} & = & \frac{1}{2} \left[ F_\mu^\alpha F_{\nu \alpha} + F_\nu^\alpha F_{\mu \alpha} - \frac{1}{2} g_{\mu \nu} F^{\alpha \beta} F_{\alpha \beta} \right] \,,
\eea

\noindent
the dark matter candidate being the (weakly coupled) lightest stable particle, which could be a scalar $X$, fermion $\chi$ or vector $X^\mu$ with corresponding field-strength tensor $F^{\mu \nu}$~\footnote{ A diagonal term proportional to the scalar potential should be present in $T^0_{\mu \nu}$ but can be safely neglected for our purposes if the typical couplings of the hidden sectors are perturbative.}. The energy--momentum tensor $T^{\mu \nu}_{\mrm{SM}}$ will be the source of production to populate the dark sector provided the interaction (\ref{eq:interaction_energymomentumtensor}).

As a first step, for simplicity, we consider only one SM scalar degree of freedom $\phi$, 
the SM Higgs\footnote{To be more precise, in the very early Universe, the electroweak symmetry is restored, and the Higgs field consists of 4 degrees of freedom. We take into account all the degrees of freedom for our analysis as discussed further on.}, whose contribution to the matter action can be written as
\beq
S_m = \int \diff^4x \sqrt{-g} {\cal L}_m
=\int \diff ^4 x \sqrt{-g} 
\left[\frac{1}{2} g_{\alpha \beta}\partial^\alpha \phi \partial^\beta \phi -V(\phi)\right]
\label{Eq:action}
\eeq

\noindent
Developing the stress-energy tensor for $\phi$ in Eq.~(\ref{eq:interaction_energymomentumtensor}) we generate an interaction term, in the case of a scalar dark matter candidate 
in the hidden sector, as
\beq
{\cal L}_\text{portal} = -\frac{1}{2} 
\left(
\frac{T_{\mu \nu}}{\Lambda^4}
\partial^\mu \phi \partial^\nu \phi 
-\frac{T^\mu_\mu}{2 \Lambda^4} \partial_\rho \phi \partial^\rho \phi
+V(\phi) \frac{T^\mu_\mu}{\Lambda^4}
\right)
\eeq
which reduces to\footnote{Notice that we neglected the scalar potential term in~(\ref{Eq:deltalm}) as well as covariant derivative for simplicity. Such terms would generate vertices involving more particles and/or perturbative couplings corresponding to processes suppressed and irrelevant for our phenomenological purposes~\cite{Giudice:1998ck}.}
\beq
{\cal L}_\text{portal} = -\frac{1}{2 \Lambda^4}
\partial_\mu X \partial^\mu \phi \partial_\nu X \partial^\nu \phi
\label{Eq:deltalm}
\eeq

\noindent
where $X$ is the lightest stable neutral scalar particle in the hidden sector. Generalization to fermionic and vectorial dark matter is straightforward.
Equation~(\ref{Eq:deltalm}) helps to understand the mechanism by which the hidden (dark) sector $X$ particle can be populated through a coupling of a form of Eq.~(\ref{eq:interaction_energymomentumtensor}), which we call {\it energy--momentum} portal. Indeed,
this term generates a 4-body interaction leading to processes such as $\phi~\phi \rightarrow X ~X$ which is largely suppressed by a factor $\sim \frac{1}{\Lambda^4}$, preventing a population of $X$ to thermalize with the SM sector in the early universe. However, even if this scattering rate is suppressed, a sufficiently
large reheating temperature of the SM thermal bath (related to the typical momentum of the particles in the plasma) can compensate
the feeble coupling required to obtain the relic abundance as observed by Planck at the present time $\Omega_X h^2 \simeq 0.1$.

In an emergent gravity scenario, effective interactions of the form of Eq.~(\ref{eq:interaction_energymomentumtensor}) may naturally arise. In this context, the deviation $h_{\mu \nu}$ of the metric $g_{\mu \nu}$ around the Minkowski metric $\eta_{\mu \nu}$ induced by the secluded sector~\cite{Kiritsis:2014yqa,BBKNpaper} can be identified as 
\beq
g_{\mu \nu} \simeq  \eta_{\mu \nu} + \frac{h_{\mu \nu}}{M_P} \equiv \eta_{\mu \nu}+ \frac{T_{\mu \nu}}{\Lambda^4} ,
\label{Eq:metric}
\eeq
\noindent
where $T_{\mu\nu}$ is a linear combination of  energy--momentum tensors of hidden sector particles. These fields generate new couplings to the SM sector through $g_{\mu \nu}$ 
\beq
S_m \rightarrow S_m + \delta S_m ~~\mrm{with}~~
\delta S_m = \int \diff ^4x \sqrt{-g} \delta {\cal L}_m,
\nonumber
\eeq
\noindent
leading to
\begin{equation}
    \delta S_\text{m}  =  \int \diff^4 x \sqrt{-g} \left( -\dfrac{1}{2 \Lambda^4} T_\text{SM}^{\mu \nu}  T_{\mu \nu}
    \right).
\end{equation}
which corresponds effectively to the interaction term of Eq. (\ref{eq:interaction_energymomentumtensor}).

Notice that apart from the coupling between the energy--momentum tensors of the hidden and the SM sectors, all gauge-invariant operators of the hidden sector couple to the SM gauge-invariant operators by irrelevant couplings\footnote{There are two exceptions, however, couplings of the SM with the hidden sector via the Higgs mass term and via the field strength of the hypercharge. The first term is related to the hierarchy problem. The second is related to anomalies, and the relevant coupling is zero since the hypercharge is massless and anomaly free.}.
These operators of the hidden sector are heavy except if symmetries protect them They can be $interpreted$ as $light$ $emergent$ $fields$, weakly coupled to the SM.
In such a way, the interaction given by Eq.~(\ref{eq:interaction_energymomentumtensor}) may arise.
Such operators include instanton densities, conserved global currents that are identified to emergent axions, and emergent gauge fields weakly coupled to the SM~\cite{AxionABBCK, globalcurrentstoU1, ABCKgraviphotons}. 

The rank of the gauge group of the hidden sector can be large, and its structure is without any other special postulates imposed on its field content. In \cite{Kiritsis:2014yqa,BBKNpaper} the authors assumed an $SU(N)$ group with $N$ very large, but we will not consider this multiplicity factor in the following for simplicity. One can however easily recover this factor by multiplying the reaction rate by a factor of $N$ and $N^2-1$ for fundamental and adjoint $X$, respectively.
It is also interesting to note that the metric defined by (\ref{Eq:metric}) corresponds to a pure disformal transformation~\cite{Bekenstein:1992pj}, whose consequences for a WIMP-type of dark matter has been studied very recently~\cite{Trojanowski:2020xza}.

Notice here that corrections to the DM abundance should be expected at next-to-leading order in the large $N$ expansion, as extra interaction terms between the SM and the dark sector suppressed by $N$ should be generated but are typically negligible for our purposes. The DM stability is automatic, for fermionic and vectorial DM candidates, as long as the energy--momentum tensor is the only portal to the SM and the DM is stable within the dark sector. In this case extra operators, suppressed by subleading powers of $N$, should not spoil the symmetry structure of the dark sector and therefore should not jeopardize the DM stability. In the case where the DM is a spin $1/2$ particle, stability within the dark sector can be justified by fermion number conservation for instance, if our DM candidate is the lightest fermion. If the DM candidate is a scalar, we could assume a discrete symmetry to prevent a cubic term or consider a vanishing vacuum expectation value to prevent the DM from decaying by mixing with the Higgs boson. Note also that in the scalar DM case, there can be $X^2|H|^2$ term with a coupling of order unity \cite{Kiritsis:2014yqa}, where $H$ is the SM Higgs field. To highlight the energy--momentum portal, however, we require such terms to be negligible by assuming, for instance, couplings between $X$ or $H$ and messengers sufficiently small.


Moreover, the kind of interaction written in Eq.~(\ref{eq:interaction_energymomentumtensor}) is also expected in the modified gravity scenarios where, in addition to the typical massless graviton, a massive spin-2 state is present in the spectrum and couples in the same way to the energy--momentum tensor of the SM and a dark sector. Such as massive spin-2 state is typically present in bimetric theories of gravity~\cite{Hassan:2011vm} or in extra-dimension constructions featuring Kaluza-Klein modes~\cite{Garny:2017kha}. Provided that the massive spin-2 mediator mass $m_2$ is larger than the maximum temperature of the universe, the scale $\Lambda$ can be identified as the geometrical average of $m_2$ and the Planck mass $\Lambda\sim \sqrt{M_P m_2}$. Dark matter phenomenology has been recently studied in this context~\cite{Garny:2017kha,Bernal:2018qlk}.

Finally, we want to stress that this type of portal is also of the same 
nature in high-scale SUSY~\cite{Benakli:2017whb}. Indeed, the minimal coupling
of a gravitino\footnote{The spin-$\frac{3}{2}$ superpartner of the graviton.} to the 
SM, whose longitudinal mode is the Goldstino denoted by $\Psi_{3/2}$, is built by first defining a vierbein~\cite{va}
\begin{equation}
e_\mu^\alpha \ = \ \delta_\mu^\alpha -\frac{i}{2 F^2}
\left( \partial_\mu \bar \Psi_{3/2} \gamma^\alpha { \Psi_{3/2}} +  \bar \Psi_{3/2} \gamma^\alpha \partial_\mu {\bar \Psi_{3/2}} \right)
\ , 
\label{va1}
\end{equation}
\noindent
with $\sqrt{F}$ being related to the SUSY breaking scale.
The couplings to matter follows the standard coupling to matter of a metric tensor built out from the vierbein 
$g_{\mu \nu} = \eta_{\alpha \beta }e_\mu^\alpha e_\nu^\beta $.
The corresponding effective operators are consequently of dimension eight and take the form: 
\begin{equation}
{\cal L}_{3/2} = \frac{i}{2F^2}( \bar \Psi_{3/2}\gamma^\mu\partial^\nu \Psi_{3/2} - \partial^\nu \bar \Psi_{3/2} \gamma^\mu\Psi_{3/2}) T^{\rm SM}_{\mu\nu}.
\end{equation}
In this supergravity context, the metric plays the same role than in 
emergent gravity scenario, as a portal whose development
generates new couplings to the SM, allowing to produce dark matter (gravitino in this case) in a sufficient amount to fulfill the Planck
constraints, as shown in~\cite{Benakli:2017whb}.

\section{III. Dark matter phenomenology}

\subsection{Production from scattering}

\begin{figure}[ht]
\centering
\includegraphics[width=2.in]{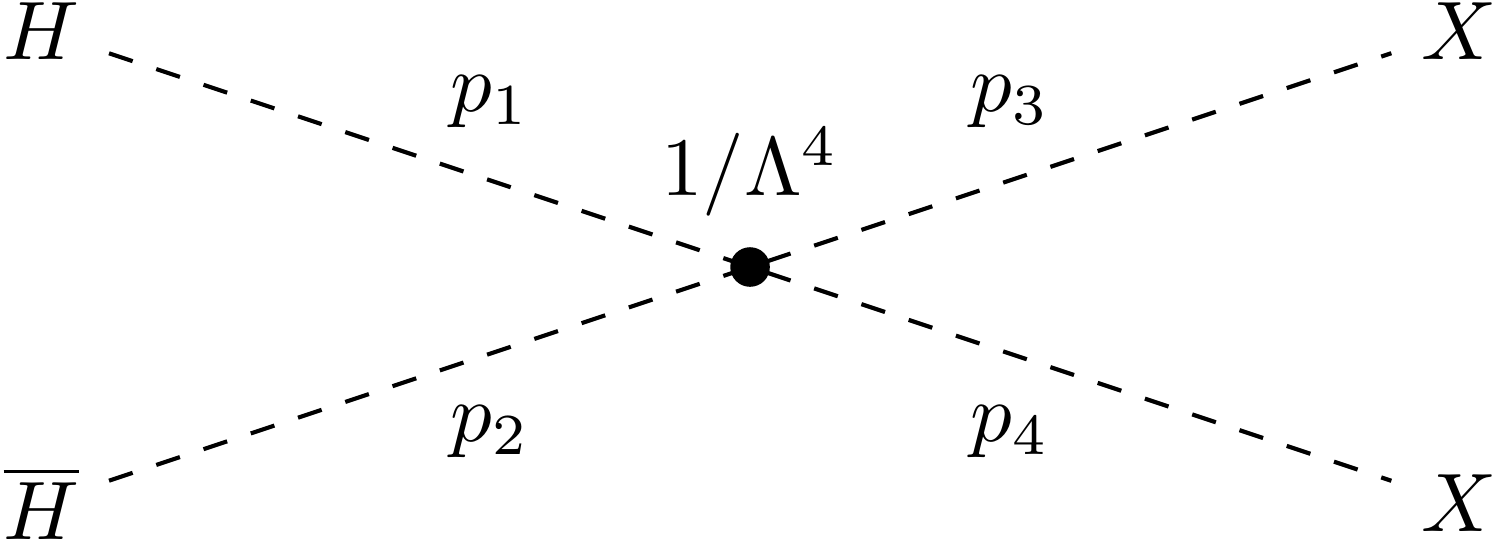}
\caption{\em \small Example of a Feynman diagram for the dark matter production. We draw the process involving the Higgs scattering in the thermal bath, whereas all the particles of the Standard Model should be considered in the complete calculation.}
\label{Fig:feynman}
\end{figure}

Once we have understood the source of dark matter, we need to compute
the relic abundance produced by the scattering of the SM particles
in the thermal bath. We will suppose that the hidden sector is secluded
from the visible sector, including the inflaton {\it i.e.} we do not consider
the possibility for the inflaton to decay directly into the hidden sector. 
However, radiative decay can be generated at the loop level, as was shown in
\cite{Kaneta:2019zgw}. We will consider this possibility in a following section.
We represent a typical process of dark matter production from scattering of scalar particles in the thermal bath in Fig.~\ref{Fig:feynman},
where we identified the SM field $\phi$ by the Higgs field. For reheating temperature $T_{RH}$ below the messenger scale $\Lambda$, our approximation in Eq.~(\ref{Eq:metric}) is justified and the coupling between the two sectors is highly suppressed by a factor $\frac{1}{\Lambda^4}$, suited for a DM population generated by the freeze-in mechanism.

To compute the dark matter number density $n_X$, one needs to solve the Boltzmann equation 
\beq
\frac{\diff n_X}{\diff t} + 3 H n_X \,=\, R(t)
\eeq
\noindent
where $R(t)$ denotes the production rate of dark matter (per unit volume per unit time). Expressing the equations as function of the thermal bath temperature, and supposing an instantaneous reheating, one can write~\footnote{we neglect the temperature evolution of the relativistic degrees of freedom for simplicity.}
\beq
\frac{\diff  Y_{X}}{\diff T}\, = \,- \frac{R(T)}{H (T)~T^4}
\label{Eq:boltzmann}
\eeq
\noindent
with $Y_{X} = \frac{n_{X}}{T^3}$, $H(T)= \sqrt{\frac{g_* \pi^2}{90}} \frac{T^2}{M_P}$, $g_*$ being the effective relativistic degrees of freedom at the temperature $T$. The production rate $R(T)$ is associated to processes such as $1+2 \rightarrow 3+4$ where 1, 2 denote particles of the SM and 3, 4 DM states as represented in Fig.~\ref{Fig:feynman} and can be schematically written as
\beq
R(T) = \frac{1}{1024 \pi^6}\int f_1 f_2 E_1 \diff E_1 E_2 \diff E_2 \diff \cos \theta_{12}\int |{\cal M}|^2 \diff \Omega_{13}.
\nonumber
\eeq
\noindent
Details regarding the computation are provided in Appendix A.
After integration, and the sum on all the degrees of freedom constituting the thermal
bath, we obtained for a dark matter particle of spin $i$
\beq
R(T) = \beta_i \frac{T^{12}}{\Lambda^8}
\eeq
\noindent
with
\bea
\beta_0\,= & \,\dfrac{121069\pi^7}{76204800} \simeq 4.8\,,
\nonumber
\\
\beta_{1/2}\,= & \,\dfrac{281287 \pi^7}{38102400}  \simeq  22.3\,,
\nonumber
\\
\beta_{1}\,= & \,\dfrac{25433 \pi ^7}{1555200} \simeq  49.4 . \nonumber
\eea

\noindent
After integration of Eq.~(\ref{Eq:boltzmann})
the density at $T_{RH}$ can be written
\beq
n_i(T_{RH}) = \frac{3 \sqrt{10}\beta_i M_P}{7 \pi \sqrt{g_{RH}}} \frac{T^{10}_{RH}}{\Lambda^8}.
\label{Eq:nDM}
\eeq
\noindent
where $g_{RH}$ is the number of degrees of freedom at the reheating time
($g_{RH} =106.75$ in the SM).
We can then deduce the relic abundance from scattering today
\beq
\Omega h^2 =1.7 \times 10^{8} 
\left(\frac{g_0}{g_{RH}} \right)
\frac{n_i(T_{RH})}{T_{RH}^3} 
\left(\frac{m_i}{1~\mrm{GeV}} \right)
\label{Eq:omega}
\eeq
\noindent
which gives
\bea
&&
\Omega h^2_{scat}\simeq 1.7 \times 10^{8} 
\left(\frac{g_0}{g_{RH}} \right)
\frac{3\sqrt{10}\beta_i M_P}{7 \pi \sqrt{g_{RH}}} \frac{T_{RH}^7}{\Lambda^8}
\left(\frac{m_i}{1~\mrm{GeV}} \right)
\nonumber
\\
&&
\simeq 0.1 \left(\frac{T_{RH}}{3 \times 10^{11} ~\mrm{GeV}} \right)^7
\left(\frac{10^{14}~\mrm{GeV}}{\Lambda} \right)^8
\left(\frac{\beta_i m_i}{0.1~\mrm{EeV}} \right)
\label{Eq:omegah2}
\eea

\noindent
where $g_0=3.91$ the present effective number of degrees of freedom and we took $g_{RH}=106.75$ corresponding to the total SM relativistic degrees of freedom.
We notice that we need very heavy dark matter, of the order of $10^{8}$ GeV
to compensate the very tiny rate of production, reduced by a factor $\frac{1}{\Lambda^8}$.

\subsection{Production from inflaton decay}

\begin{figure}[ht]
\centering
\includegraphics[width=2.5in]{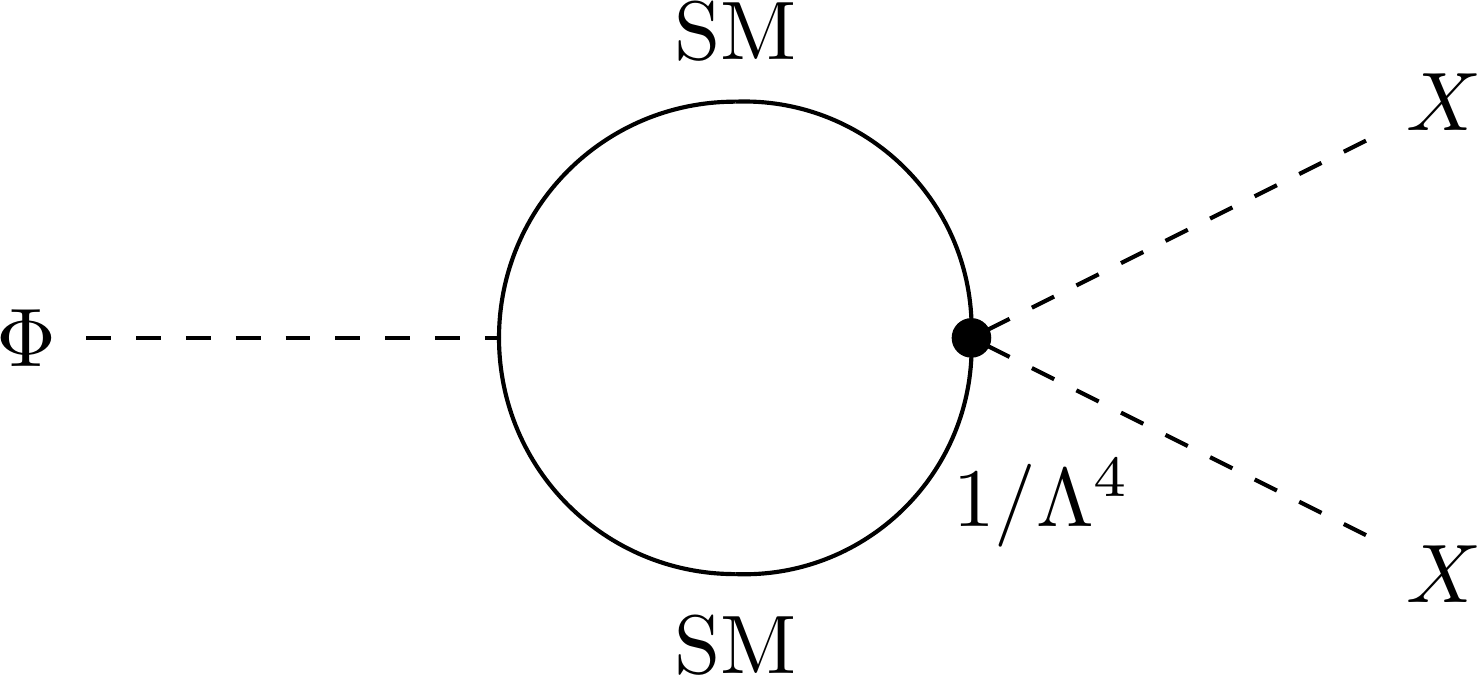}
\caption{\em \small Production of dark matter via loop-induced inflaton decay}
\label{Fig:feynmanbis}
\end{figure}

It was shown in~\cite{Kaneta:2019zgw} that if dark matter is produced by 
scattering one cannot avoid the direct production of dark matter through the loop-induced inflaton decay, as depicted in Fig.~\ref{Fig:feynmanbis}.
The minimal way to couple the Standard Model sector
to the inflaton field is through the Higgs boson:
\beq
{\cal L}_\Phi = \mu_\Phi \Phi|H|^2
\label{Eq:inflaton-higgs}
\eeq
\noindent
If one consider only the Higgs running in the loop of Fig.~\ref{Fig:feynmanbis},
we obtain

\bea
&&
\Gamma^\Phi_{XX} = 
\frac{|\frac{2}{3}-i \pi|^2}{72 \pi (16 \pi^2)^2}
\frac{\mu_\Phi^2 m_\Phi^7}{\Lambda^8}
\label{Eq:widths}
\eea

\noindent 
where we have supposed a scalar dark matter for illustration. On the other hand, the main perturbative decay width of the inflaton which will generate the thermal bath in the reheating stage is given by
\bea
\Gamma^\Phi_{HH}= \frac{\mu_\Phi^2}{8 \pi m_\Phi}.
\nonumber
\eea
\noindent
Remark that the coupling (\ref{Eq:inflaton-higgs}) also induces an additional term given as ${\cal L}\supset (\mu_\Phi/2\Lambda^4)\Phi|H|^2 T^\mu_\mu$, since the  energy-momentum tensor for the Higgs boson acquires the term $-g_{\mu\nu}{\cal L}_\Phi$.
As a result, a 4-body decay channel, $\Phi\to XXHH$ is allowed.
However, it turns out that this process is subdominant compared to the loop induced decay.
In addition, the same five-point vertex gives additional contribution for $\Phi\to HH$ through the Higgs loop, which is however suppressed by the Higgs mass $m_h$, as the effective coupling takes the form of $(\mu_\Phi m_h^2/\Lambda^4)\Phi (\partial^\mu X\partial_\mu X)$. 

Results for
fermionic and vectorial dark matter are given 
in Appendix B. Notice that the coupling of the inflaton 
field to the SM, $\mu_\Phi$, {\it does not} 
appear in the branching ratio\footnote{That would be the case, whatever is the effective coupling one would consider between the inflaton and the Standard Model particles, e.g.,

\beq
B_R = \frac{N_X\Gamma^\Phi_{XX}}{\Gamma^\Phi_{HH}} \simeq
\frac{|\frac{2}{3} - i \pi|^2}{9(16 \pi^2)^2} \frac{m_\Phi^8}{\Lambda^8},
\eeq

\noindent
where $N_X$ is the number of $X$ particles produced per decay, which is $N_X=2$ in the present case.
}.
Supposing an instantaneous reheating, we can write
\bea
&&
n_{dec}(T_{RH})= B_R \frac{\rho_\Phi(T_{RH})}{m_\Phi}
=B_R \frac{\rho_R(T_{RH})}{m_\Phi} 
\nonumber
\\
&&
= B_R
\left(\frac{g_{RH} \pi^2}{30} \right)
\frac{T_{RH}^4}{m_\Phi}
\nonumber
\eea

\noindent
where we have {\it defined} the reheating temperature by the condition $\rho_\Phi(T_{RH})$ = $\rho_R(T_{RH})$, in other words, when radiation and inflaton densities equilibrate. Notice that different definitions of the reheating temperature can lead to slightly different results, but differing never more than factors of the order of unity. 
After combining with Eq.~(\ref{Eq:omega}), we obtain
\bea
&&
\Omega h^2_{dec} = 0.17 \frac{B_R}{ 10^{-9}}
\left(\frac{g_0 \pi^2}{30} \right)
\left( \frac{T_{RH}}{m_\Phi}\right)
\left( \frac{m_X}{1~\mrm{GeV}} \right)
\label{Eq:omegadecayinstant}
\\
&&
=0.17\frac{\frac{4}{9} + \pi^2}{9\times 10^{-9}(16 \pi^2)^2} \left( \frac{T_{RH}m_\Phi^7}{\Lambda^8} \right)
\left( \frac{m_X}{1~\mrm{GeV}} \right)
\nonumber
\\
&&
\simeq 0.17 \left( \frac{T_{RH}}{10^{10}}\right)
\left( \frac{m_\Phi}{3 \times 10^{13}} \right)^7
\left(\frac{10^{14}}{\Lambda} \right)^8
\left( \frac{m_X}{1~\mrm{TeV}} \right).
\nonumber
\eea

\noindent
where the units are in GeV when not specified.
When we compare Eq.~(\ref{Eq:omegadecayinstant}) with Eq.~(\ref{Eq:omegah2}), we
see that the dependence on the reheating temperature for the
production of dark matter through the inflaton decay is much 
weaker than for the production process through scattering.
As a consequence, for light dark matter, where one needs
a large amount of dark matter to fulfill Planck constraints, 
the scattering process will be the one able to populate the Universe. On the other hand, for heavier dark matter, the bound will be quickly saturated by the decay process.

As an exercise, we can compute the mass of dark matter for which
both sources, scattering and decay, equalize:
\beq
\Omega h^2_{scat}= \Omega h^2_{dec} 
~~\Rightarrow ~~
T_{RH}^{eq} \simeq 2.6 \times 10^{12} ~\mrm{GeV}.
\eeq
\noindent 
for a scalar dark matter\footnote{The result differs just by a factor $\beta_i$ for vectorial dark matter.}. Remarkably,
the reheating temperature for which the transition between 
a regime dominated by scattering production to a regime
dominated by decay production does not depend on the dark matter mass. For $T_{RH} > T_{RH}^{eq}$, the scattering process populate the dark sector, whereas for $T_{RH} < T_{RH}^{eq}$, the inflaton decay is the main source for the relic abundance.

\begin{figure}[ht]
\centering
\includegraphics[width=3.in]{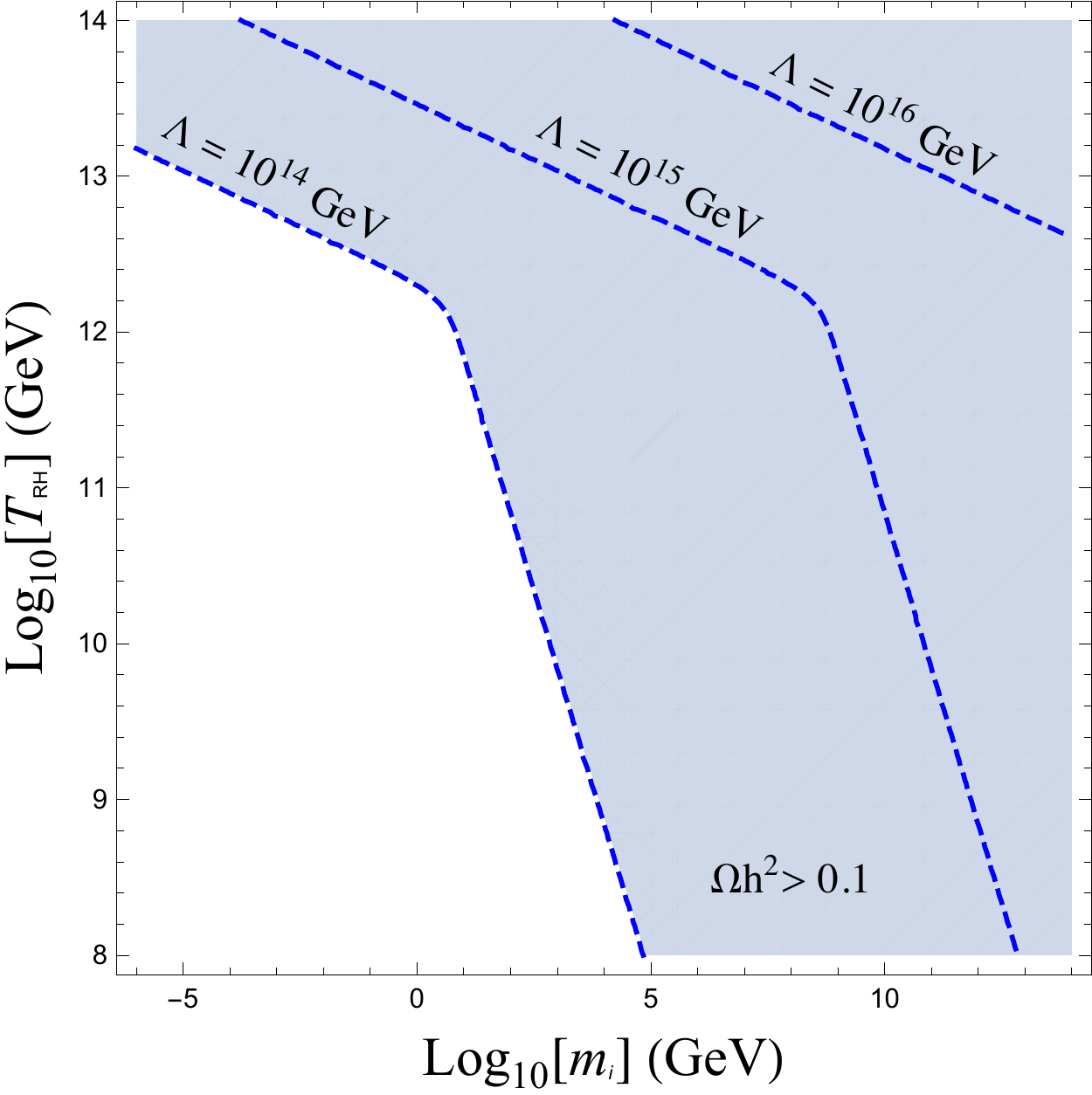}
\caption{\em \small Parameter space allowed in the ($m_i$, $T_{RH}$)
parameter space, for different values of $\Lambda$ for a scalar dark matter.}
\label{Fig:omega0}
\end{figure}

\noindent
We show in Fig.~\ref{Fig:omega0} the parameter space respecting the correct relic abundance in the case of a scalar dark matter. We recognize clearly the two regimes (scattering on the left and decay on the right) from
their different dependence on the reheating temperature, especially the change of regime for $T_{RH}=T_{RH}^{eq} \simeq 2\times 10^{12}$ GeV.

\subsection{Non-instantaneous reheating effects}
\label{sec:non-inst}

Several recent works showed that the naive instantaneous approximation is not valid anymore
if one has to deal with highly temperature-dependent production processes. Effects of non-instantaneous reheating were studied in~\cite{Garcia:2017tuj} whereas non-instantaneous thermalization in~\cite{Reheating} and effects of the inflaton potential in~\cite{Garcia:2020eof}. The effects begin to be important when, if we write the production rate $R(T) \propto \frac{T^{n+6}}{\Lambda^{n+2}}$, values of
$n\geq 6$, which is precisely our case.
We discuss in this section the dark matter production by incorporating non-instantaneous reheating effects.
In the following we assume by simplicity that the thermalization completes instantaneously, and the preheating stage is negligible.
We also assume that after the end of inflation, the inflaton oscillation is described by a quadratic potential, given by $V(\Phi)=\frac{1}{2} m_\Phi^2\Phi^2$.




The reheating is generated by the perturbative inflaton decay which can be parametrized by
\bea
\Gamma^\Phi &=& \frac{y^2}{8\pi}m_\Phi.
\eea
\noindent
It is a commonly used generic form for the width
of the inflaton, and $y \equiv \mu_\Phi/m_\Phi$ 
when the decay is dominated by the Higgses channel (\ref{Eq:inflaton-higgs}). The non-instantaneous reheating is characterized by a maximum temperature, $T_{\rm max} > T_{RH}$, where the production rate of the dark matter is more important.
The maximal temperature $T_{\rm max}$ and $T_{RH}$ can be computed solving the set of Friedmann equations
\bea
&&
\frac{\diff\rho_\Phi}{\diff t}+3H\rho_\Phi=-\Gamma^\Phi\rho_\Phi,
\label{Eq:noninst1}
\\
&&
\frac{\diff\rho_R}{\diff t}+4H\rho_R=+\Gamma^\Phi\rho_\Phi,
\label{Eq:noninst2}
\\
&&
H^2= \frac{\rho_\Phi+\rho_R}{3 M_P^2}
\label{Eq:noninst3}
\eea

\noindent
The solution can be found in~\cite{Garcia:2020eof} for a generic type of inflaton potential. They obtained
\bea
T_{\rm max} &=& \left(\frac{45}{32}\frac{3^{1/10}}{2^{4/5}}\frac{y^2 m_\Phi M_P \rho_{\rm end}^{1/2}}{g_*(T_{\rm max}) \pi^3}\right)^{1/4}\\
&\simeq& 8.8\times10^{14}~{\rm GeV}\times y^{1/2}\left(\frac{106.75}{g_*(T_{\rm max})}\right)^{1/4}\nonumber\\
&&\times \left(\frac{m_\Phi}{3\times10^{13}~{\rm GeV}}\right)^{1/4}\left(\frac{\rho_{\rm end}}{0.175m_\Phi^2M_P^2}\right)^{1/8},\nonumber\\
T_{RH} &=& \left(\frac{9}{40}\frac{y^4 m_\Phi^2 M_P^2}{g_{RH}\pi^4}\right)^{1/4}\\
&\simeq&5.8\times10^{14}~{\rm GeV}\nonumber\\
&&\times y\left(\frac{106.75}{g_{RH}}\right)^{1/4}\left(\frac{m_\Phi}{3\times10^{13}~{\rm GeV}}\right)^{1/2}\nonumber
\eea

\noindent
where $T_{RH}$ is defined as the temperature where the energy densities of the inflaton $(\rho_\Phi)$ and the radiation $(\rho_R)$ become equal.
In the following we will take $g_*(T_{\rm max})=g_{RH}$. $\rho_{\rm end}$ is the inflaton energy density at the end of inflation, which can be computed by requiring that the equation-of-state parameter $w$ reaches the value $w=-1/3$ at which the scale factor $a$ satisfies $\diff ^2a/\diff t^2=0$.
For instance, $\rho_{\rm end}\simeq 0.175 m_\Phi^2 M_P^2$ for the Starobinsky-type of inflaton potential (\cite{Ellis:2015pla}, see the Appendix C for details), which will be used as a concrete example in the following analysis.
If one takes into account the $T_\text{max}$ effect induced by the non-instantaneous reheating, the relic abundance is increased by a
{\it boost factor} $B^{\rm scatt}\equiv n_i^{\rm non-inst}(T_{RH})/n_i(T_{RH})$ which was calculated in~\cite{Garcia:2017tuj} :
\bea
B^{\rm scatt}=c\frac{56}{3}\log\frac{T_{\rm max}}{T_{RH}},
\eea
where a numerical factor $c\simeq 0.45$ arises to match the numerical results where the reheating temperature shifts slightly toward higher values than the pure analytical one. To compute the relic abundance, one has to add to the system of equations (\ref{Eq:noninst1}-\ref{Eq:noninst3}), the Boltzmann equation for the dark matter  

\bea
\frac{\diff n_X}{\diff t}+3Hn_X&=&R(t)+ B_R\Gamma^\Phi\rho_\Phi/m_\Phi,
\label{Eq:noninst4}
\eea

\noindent
with the initial conditions $\rho_\Phi(t=0)=\rho_{\rm end}, \rho_R(t=0)=n_X(t=0)=0$.
As we consider the Starobinsky model for inflation, the coupling $y$ is restricted by the inflationary predictions, especially the spectral index $n_s$.
A detailed discussion can be found in Appendix C, and in our case, $y$ should satisfy $y\gtrsim3.7\times10^{-6}$.

\begin{figure*}[t]
\centering
\includegraphics[width=7.2in]{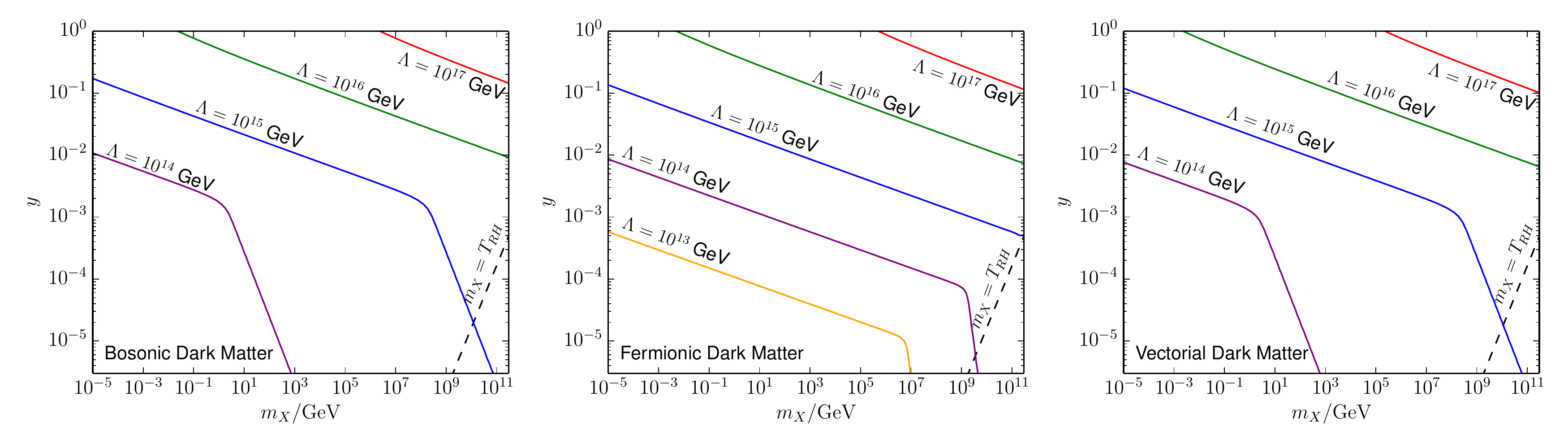}
\caption{\em \small Relic abundance constraint for spin $0$, $\frac{1}{2}$ and $1$ dark matter in the case of non-instantaneous reheating for different values of the messenger scale $\Lambda$. See the text for details.
}
\label{Fig:Omega_noninst}
\end{figure*}

\noindent
We show in Fig.~\ref{Fig:Omega_noninst} the result of our numerical integration of the full system of 4 equations, in 
the ($m_i$, $y$), for different values of $\Lambda$. The line correspond to points respecting the Planck result, for bosonic, fermionic and vectorial dark matter.
As we expected, the parameter space allowed lighter dark matter, or in another words, larger values of $\Lambda$ for a given dark matter mass. This comes from the fact that the boost factor enhanced the production mechanism. One then necessitates lighter dark matter or larger BSM scale to avoid overabundance. 

We also remark that the nature of the dark matter do not play an important role in the scattering production regime, as their production rate is similar up to numerical factors of order of unity.
On the other hand, in the decay production regime, the viable parameter space for fermionic dark matter may be distinguishable from the other two, since the inflaton decay into a pair of fermionic dark matter is suppressed by the dark matter mass. This can be seen comparing Eqs.(\ref{Eq:widthscalar}) and (\ref{Eq:widthfermion}), the mass dependence coming from the presence of a $\gamma^\mu$ matrix in the coupling
(see also Appendix B).
Notice also that for small values of $y$ and/or large $m_i$, there is a region where the DM mass becomes larger than $T_{RH}$.
In this case an additional entropy production dilutes the DM number density, which we have incorporated in the figure.



We also want to underline that
in the case of non-instantaneous reheating, the contribution coming from the inflaton decay is just mildly different from the instantaneous case, and can be written~\cite{Kaneta:2019zgw}
\bea
&&
\Omega h^2_{dec} = \frac{B_R}{2.9 \times 10^{-9}} 
\left( \frac{T_{RH}}{m_\Phi}\right)
\left( \frac{m_X}{1~\mrm{GeV}} \right)
\label{Eq:omegadecay}
\\
&&
=\frac{\frac{4}{9} + \pi^2}{2.6\times 10^{-8}(16 \pi^2)^2} \left( \frac{T_{RH}m_\Phi^7}{\Lambda^8} \right)
\left( \frac{m_X}{1~\mrm{GeV}} \right)
\nonumber
\\
&&
\simeq 0.1 \left( \frac{T_{RH}}{10^{10}}\right)
\left( \frac{m_\Phi}{3 \times 10^{13}} \right)^7
\left(\frac{10^{14}}{\Lambda} \right)^8
\left( \frac{m_X}{144~\mrm{GeV}} \right).
\nonumber
\eea

\noindent

Before closing the section, we briefly comment on the assumption of the instantaneous thermalization which we have made in our analysis.
In general the thermalization is also not an instantaneous process.
The time scale for the thermalization to complete, however, depends on the non-thermal population of the SM particles produced by the inflaton decay.
In our particular case based on the Starobinsky model, the effect of the non-instantaneous thermalization becomes non-negligible for $y\lesssim 10^{-5}$ \cite{Ellis:2015jpg}.
In such regions we may expect another enhancement in the DM production \cite{Reheating}, though a further discussion on this effect is beyond the scope of this paper.

\section{IV. Conclusions}

We have shown that models where gravity emerge 
from a secluded hidden sector, can contain a viable dark matter
candidate. Its production can occurs by the scattering of the SM particles in the thermal bath, or by the inflaton decay. We have computed the relic abundance of dark matter
as a function of the reheating temperature for different values
of the BSM scale $\Lambda$, representing the mass of the heavy
messengers secluding the two sectors. We have shown that for reasonable values of $T_{RH}\simeq 10^{10}$ GeV and dark matter masses $m_X \lesssim m_\Phi\simeq 3\times10^{13}$ GeV, a large part of the parameter
space satisfies Planck constraints. We studied the case of scalar, fermionic and vectorial dark matter, with similar conclusions.
Moreover, our framework is much more general than emergent gravity--like constructions. Indeed, the fundamental ingredient
for the dark matter production lies in terms of the type
$\frac{1}{\Lambda^4} T^{\mu \nu}_X T_{\mu \nu}^{SM}$,
that we recover, for instance, in minimal supergravity scenarios.
We suggest that this novel type of portal,
that we called {\it Energy-Momentum portal}, 
should be present in a large class of UV models where the 
secluded sector can mix with kinetic terms of the Standard Model.

\vskip.1in
{\bf Acknowledgments:}
\noindent 
The authors want to thank especially P. Brax, E. Dudas, E. Kiritsis and P. Betzios for very  insightful
discussions. This work was supported in part by the France-US PICS MicroDark and the ANR grant Black-dS-String ANR-16-CE31-0004-01. MP would also like to thank the
Paris-Saclay Particle Symposium 2019 with the support of the P2I and SPU research departments and
the P2IO Laboratory of Excellence (program "Investissements d'avenir"
ANR-11-IDEX-0003-01 Paris-Saclay and ANR-10-LABX-0038), as well as the
IPhT. The work of MP was supported by the Spanish Agencia Estatal de Investigaci\'{o}n through the grants FPA2015-65929-P (MINECO/FEDER, UE),  PGC2018-095161-B-I00, IFT Centro de Excelencia Severo Ochoa SEV-2016-0597, and Red Consolider MultiDark FPA2017-90566-REDC. 
This project has received funding/support from the European Unions Horizon 2020 research and
innovation programme under the Marie Skodowska-Curie grant agreements Elusives ITN No. 674896
and InvisiblesPlus RISE No. 690575. 
The work of KK was supported in part by the DOE grant DE-SC0011842 at the University of Minnesota.
P.A. was supported by FWF Austrian Science Fund via the SAP P30531-N27.

\section*{Appendix}

\section{A. Production rate:  scattering}
\label{sec:productionrate}

The Boltzmann equation for the DM number density can be written as
\begin{align}
    \dfrac{\diff n_\text{DM}}{\diff t}+3Hn_\text{DM}\,= \,R(T)
\end{align}
where the quantity on the right-hand-side $R(T)$ represents the temperature-dependent DM production rate per unit of volume and time which can be expressed as
\begin{widetext}
\begin{equation}
  R(T)  \,\equiv \,\sum_{1,2\in\text{SM}}\,2\,\left( \dfrac{1}{1+\delta_{12}}\right)\,\int \diff \Pi_1 \diff \Pi_2 \diff \Pi_3 \diff \Pi_4 f_1 (p_1) f_2 (p_2) (2\pi)^4 \delta^{(4)}(p_1+p_2-p_3-p_4) |\mathcal{M}|^2
  \label{eq:defRate}
\end{equation}
\end{widetext}
where
\begin{equation}
    \diff \Pi_i \equiv \dfrac{\diff^3 p_i}{(2\pi)^3}\dfrac{1}{2 E_i}
\end{equation}
with $p_i(E_i)$ being the 4-momentum (energy) of particle $i$.  It is the sum over all processes $1+2\rightarrow 3+4$ with 1, 2 being particles of the SM and 3, 4 dark matter states.  The factor of 2 account for the production of 2 DM particle per process and the factor $(1+\delta_{12})^{-1}$ prevents over-counting the phase space volume for identical initial states. $|\mathcal{M}|^2$ is the matrix element squared summed over polarization states and divided by symmetry factor for identical particles in the final state. $f_{1,2}$ are the phase space distribution functions of the initial state. As the amplitude squared depends only on the spin of initial and final state, for a dark matter candidate of a given spin, the production rate can be expressed as a sum over all possible initial states present in the SM with a spin $j=0,1/2,1$ :
\begin{widetext}
\begin{equation}
     R(T)\,=\,\sum_{j=0,1/2,1} N_j R_j\,=\,
    N_j \,  \frac{1}{1024 \pi^6}\int f_j(E_1) f_j(E_2) E_1 \diff E_1 E_2 \diff E_2 \diff \cos \theta_{12}\int |{\cal M}_j|^2 \diff \Omega_{13} \, = \,  4R_0+45R_{1/2}+12R_1 
\end{equation}
\end{widetext}
in the limit where all species are relativistic. $N_j$ denotes the number of each SM species of spin $j$ : $N_0=4$ real scalar degrees of freedom (1 complex Higgs doublet), $N_1=12$ gauge bosons (8 gluons, 4 electroweak bosons) and $N_{1/2}=45$ Dirac fermions (6 quarks with 3 colors + anti-particles, 3 charged leptons + antiparticles, 3 neutrinos). $R_j$ denotes the contribution to the rate of one pair of particle 1, 2 with spin $j$. $f_j$ is the Fermi-Dirac $(j=1/2)$ or Bose-Einstein ($j=0,1$) distribution. The infinitesimal solid angle is defined as
\begin{equation}
    \diff \Omega_{13}=2 \pi \diff \cos \theta_{13}
\end{equation}
with $\theta_{13}$ and $\theta_{12}$ being the angle formed by momenta of 1,3 and 1,2 respectively. In the massless limit, the amplitude squared can be expressed in terms of Mandelstam variable $s$ and $t$, related to the angle $\theta_{13}$ and $\theta_{12}$ by the relations
\begin{align}
    t\,= & \, \dfrac{s}{2}( \cos \theta_{13}-1) \\  s\, = & \,2E_1E_2(1-\cos \theta_{12})
\end{align}

The amplitudes and rates for scalar, fermionic and vector dark matter are detailed in the following subsections. 

\subsection{Scalar dark matter}
\begin{equation}
 |\mathcal{M}_0|^2=\dfrac{(s^2+2st+2t^2)^2}
 {32 \Lambda^8}
\end{equation}
\begin{equation}
 |\mathcal{M}_{1/2}|^2=-\dfrac{t(s+t)(s+2t)^2}
 {16 \Lambda^8}
\end{equation}
\begin{equation}
 |\mathcal{M}_{1}|^2=\dfrac{t^2(s+t)^2}
 {4 \Lambda^8}
\end{equation}
The total rate is given by
\begin{equation}
    R(T)= \dfrac{121069\pi^7}{76204800} \dfrac{T^{12}}{\Lambda^8} \equiv \beta_0 \dfrac{T^{12}}{\Lambda^8} 
\end{equation}
with $\beta_0 \simeq 4.8$.

\subsection{Fermionic dark matter}
\begin{equation}
 |\mathcal{M}_0|^2=-\dfrac{t(s+t)(s+2t)^2}
 {8 \Lambda^8}
\end{equation}
\begin{equation}
 |\mathcal{M}_{1/2}|^2=\frac{s^4+10 s^3 t+42 s^2 t^2+64 s t^3+32 t^4}{32 \Lambda ^8}
\end{equation}
\begin{equation}
 |\mathcal{M}_{1}|^2=-\frac{t (s+t) \left(s^2+2 s t+2 t^2\right)}{2 \Lambda^8}
\end{equation}
The total rate is given by
\begin{equation}
    R(T)=\frac{281287 \pi^7}{38102400} \dfrac{T^{12}}{\Lambda^8} \equiv \beta_{1/2} \dfrac{T^{12}}{\Lambda^8} 
\end{equation}
with $\beta_{1/2} \simeq 22.3$.

\subsection{Vector dark matter}
\begin{equation}
 |\mathcal{M}_0|^2=\frac{s^4+4 s^3 t+16 s^2 t^2+24 s t^3+12 t^4}{32 \Lambda ^8}
\end{equation}
\begin{equation}
 |\mathcal{M}_{1/2}|^2=-\frac{t (s+t) \left(5s^2+12 s t+12 t^2\right)}{16\Lambda^8}
\end{equation}
\begin{equation}
 |\mathcal{M}_{1}|^2=\frac{\left(s^2+s t+t^2\right) \left(s^2+3 s t+3 t^2\right)}{4 \Lambda ^8}
\end{equation}
The total rate is given by
\begin{equation}
    R(T)=\frac{25433 \pi ^7}{1555200}\dfrac{T^{12}}{\Lambda^8} \equiv \beta_{1} \dfrac{T^{12}}{\Lambda^8} 
\end{equation}
with $\beta_{1} \simeq 49.4$.

\section{B. Production rate: decay}
\label{Sec:appendixb}

This appendix contains a complete list of the production rate of radiative inflaton decays.
As a counterpart to $R(t)$ in the scattering case, we may use $B_R\Gamma^\Phi\rho_\Phi/m_\Phi$ for the decay contribution in the Boltzmann equation.

In general, when a loop-induced inflaton coupling to dark matter is renormalizable, its effective coupling depends on the renormalization scale which we set the inflaton mass throughout our analysis.
On the other hand, a loop induced coupling of inflaton to vectorial dark matter is essentially a dimension five operator, $\Phi F^{\mu\nu}F_{\mu\nu}$, and thus, its effective coupling is finite, which means no renormalization scale dependence.

\subsection{Scalar dark matter}
The dominant decay channel of inflaton is into a pair of the Standard Model Higgs bosons, whose decay width is given as
\begin{equation}
    \Gamma^\Phi_{HH}\,=\,4 \Gamma^\Phi_{h_ih_i} \, = \, \dfrac{1}{8 \pi} \dfrac{\mu_\phi^2}{m_\phi}
\end{equation}
where $h_i$ represent the 4 real scalar degrees of freedom of the Higgs doublet. The loop-induced inflaton decay to a pair of DM particles, corresponding to the diagram of Fig.~\ref{Fig:feynmanbis} where $h_i$ is running in the loop, is given by

\bea
\Gamma^\Phi_{XX} &=& \frac{1}{72\pi(16\pi^2)^2}\left(\frac{4}{9}+\pi^2\right)\frac{\mu_\Phi^2m_\Phi^7}{\Lambda^8},
\label{Eq:widthscalar}
\eea
yielding the branching ratio

\beq
B_R = 
\frac{1}{9(16 \pi^2)^2}\left(\frac{4}{9}+\pi^2\right) \frac{m_\Phi^8}{\Lambda^8}.
\eeq

\subsection{Fermionic dark matter}
The inflaton decay into a pair of fermionic dark matter is suppressed by the dark matter mass $m_X$ due to the angular momentum conservation.
Indeed, because of the structure of the energy-momentum tensor, a loop induced effective coupling of inflaton to fermionic dark matter necessarily involves kinetic term for $X$, resulting in an $m_X$ 
direct dependence from the equation of motion.
The decay width is then given by

\bea
\Gamma^\Phi_{\bar \chi \chi} &=& \frac{1}{2\pi(16\pi^2)^2}\left(\frac{25}{81}+\frac{\pi^2}{9}\right)\frac{m_X^2\mu_\Phi^2m_\Phi^5}{\Lambda^8}.
\label{Eq:widthfermion}
\eea

\noindent
Thus, the branching ratio can be obtained as

\bea
B_R &=& \frac{8}{(16\pi^2)^2}\left(\frac{25}{81}+\frac{\pi^2}{9}\right)\frac{m_X^2m_\Phi^6}{\Lambda^8}.
\eea

\subsection{Vector dark matter}

For $m_X\ll m_\Phi$, we may use the equivalence theorem, and thus the total squared amplitude is obtained by summing up the squared amplitudes for the massless vector boson and the real scalar boson final states.
The decay width is given by

\bea
\Gamma^\Phi_{XX} &=& \frac{1}{648\pi(16\pi^2)^2}\left(22+9\pi^2\right)\frac{\mu_\Phi^2m_\Phi^7}{\Lambda^8},
\eea

\noindent
and thus the branching ratio becomes

\bea
B_R &=& \frac{1}{81(16\pi^2)^2}\left(22+9\pi^2\right)\left(\frac{m_\Phi}{\Lambda}\right)^8.
\eea

\section{C. Inflation sector}
\label{Sec:appendixc}

In section III, we have shown the non-instantaneous reheating effects on the DM production by taking the Starobinsky-type of inflaton potential \cite{Starobinsky:1980te} as a concrete example.
The inflaton potential is given as
\bea
V=\frac{3}{4}m_\Phi^2M_P^2\left(1-e^{-\sqrt{\frac{2}{3}}\Phi/M_P}\right)^2,
\eea
where the inflaton mass $m_\Phi$ is determined to match the measured curvature power spectrum $\ln(10^{10}A_{S*})=3.044$ at the pivot scale $k_*=0.05~{\rm Mpc}^{-1}$~\cite{planck,Akrami:2018odb}.
By defining $N_*$ as a number of e-folds at which the scale of $k_*$ exits the horizon, we may write
\bea
m_\Phi^2\simeq\frac{24\pi^2A_{S*}}{N_*^2}.
\eea

By numerically solving the relation
\bea
N_* &=& \ln\left[\frac{1}{\sqrt3}\left(\frac{43}{11}\right)^{1/3}\left(\frac{\pi^2}{30}\right)^{1/4}\frac{T_0}{H_0}\right]-\ln\frac{k_*}{H_0}\nonumber\\
&&-\frac{1-w_{\rm int}}{12(1+w_{\rm int})}\ln\frac{\rho_{\rm end}}{\rho_{RH}}+\frac{1}{4}\ln\frac{V_*^2}{M_P^4\rho_{\rm end}}\nonumber\\
&&-\frac{1}{12}\ln g_{RH},
\eea
we obtain $N_*$, where we assume $V_*\simeq (3/4)m_\Phi^2M_P^2$~\cite{Martin:2010kz,Liddle:2003as}.
The equation-of-state parameter $w_{\rm int}$ is defined as an e-fold averaged value during the reheating, which we take $w_{\rm int}=0$ for simplicity. 
The present day Hubble parameter ($H_0$) and temperature ($T_0$) we use are $H_0=67.36$ km/s/Mpc and $T_0=$2.7255 K.
Then, we find that the solution for $N_*$ can be fitted by $N_*\simeq55+0.33\ln y$.
From $n_s\simeq1-2/N_*$ and $n_s=0.9649\pm0.0042$ at 68 \% CL~\cite{Akrami:2018odb}, we find $y\gtrsim3.7\times10^{-6}$.

\vspace{-.5cm}
\bibliographystyle{apsrev4-1}

\end{document}